\documentstyle[12pt]{article}
\begin{document}
\begin{center}
{\large \bf 
Fractional Dynamical Behavior in Quantum Brownian Motion
}\\

\vspace*{.5in}

\normalsize 
Kyungsik Kim$^{a,*}$, Y. S. Kong$^{b}$, M. K. Yum$^{c}$  and J. T. Kim$^{d}$ \\

\vspace*{.2in}

{\em 
$^{a}$$Department$ $of$ $Physics$, $Pukyong$ $National$ $University$,\\
$Pusan$ $608$-$737$, $Korea$\\
$^{b}$$School$ $of$ $Ocean$ $Engineering$, $Pukyong$ $National$ $University$,\\
$Pusan$ $608$-$737$, $Korea$}\\
$^{c}$$Department$ $of$ $Pediatric$ $Cardiology$, $Hanyang$ $University$,\\
$Kuri$ $471-701$, $Korea$\\
$^{d}$$Department$ $of$ $Photonic$ $Engineering$, $Chosun$ $University$,\\
$Gwangju$ $501-759$, $Korea$ 

\hfill\\
\end{center} 
%
 
%
%
%
\baselineskip 24pt
\begin{center}
{\bf Abstract}
\end{center} 

\noindent
The dynamical behavior for a quantum Brownian particle is investigated 
under a random potential of the fractional iterative map on a one-dimensional lattice.
For our case, the quantum expectation values can be obtained numerically from the wave function of
the fractional Schr$\ddot{o}$dinger equation.
Particularly, the square of mean displacement which is ensemble-averaged
over our configuration
is found to be proportional approximately to $t^{\delta}$ in the long time limit, where $\delta$ $=$ $0.96 \pm 0.02$.
The power-law behavior with scaling exponents $\epsilon$ $=$ $0.98 \pm 0.02$ and $\theta$ $=$ $ 0.51 \pm 0.01$
is estimated for $ \overline {{\langle p(t) \rangle}^2}$ and 
$ \overline {{\langle f(t) \rangle}^2}$,
and the result presented is compared with other numerical calculations.
\vskip 10mm
\noindent
$PACS$: 0530, 0540\\
$Keywords$: Quantum Brownian motion; Fractional iterative map; Quantum expectation value;
Scaling exponent                                                         
\vskip 10mm
\noindent
$^{*}$Corresponding author. Tel.: +82-51-620-6354; fax: +82-51-611-6357.\\
$E-mail$ $address$: kskim@pknu.ac.kr (Kyungsik Kim ).

\newpage


\noindent
Until now, 
The Langevin equation has been played a important role in stochastic process,
classically and quantum-mechanically. 
The classical Brownian motion has mainly been represented from the Langevin equation 
among the stochastic process problems $[1]$.
The solution of this motion has analytically been obtained
under the influence of a time-dependent fluctuation and known to be 
different from that of other motions which the stable probability distribution has usually the
well-known Gaussian form. Furthermore, the natural generalization of the Brownian motion and
the Wiener stochastic process $[2]$ has been founded on the theory of stable probability 
distributions developed 
by L$\acute e$vy.

On the other hand, the random motion of the Brownian particle has generally been described as the 
fluctuation-dissipation theorem $[3]$.
In particular, the generalized Brownian motion $[4,5]$ has been showed to be equal to the Lioville equation
that is presented from the dynamical variables in the phase space.  
The generalized Brownian motion has essentially been a determined equation which has more informations 
for the time-dependence of many-body system. As a counterpart, the quantum Brownian motion  $[6,7]$
has been defined as the dynamics of a quantum particle under the the time-dependent potential
and treated extensively with the transport of quantum excitation, the directed polymer problem,
and quantum tunneling phenomena $[8-10]$. 

Recently, there have been investigated many diffusion processes
deterministically by periodic iterated maps 
on the basis of the theory of dynamical systems.
The nature of these
diffusion motions is charaterized by the temporal scaling of the
mean-square displacement $\sigma^{2}(t)\sim t^{\nu}$. For the normal diffusion
$\nu=1$, whereas in the subdiffusion $\nu<1$, and in the superdiffusion $\nu>1$.
The subdiffusion has been arised typically in amorphous semiconductor $[11]$ 
and polymer networks $[12]$ and in porous media and fractal lattices $[13]$.
The superdiffusion has been observed in rotating laminar fluid flows $[14]$ and in enhanced
transport of particles $[15]$. 
In particular, for the Kim-Kong map, the mean-square displacement for the resultant diffusive 
motion is found to scale approximately linearly with time for typical two 
control parameters $[16]$.

In reality, the dynamical behavior of a quantum particle for the quantum Brownian motion 
has analytically and numerically been discussed from the square of the mean displacement $[9,10,17]$ 
that is followed the scaling behavior of the width of the wave function, 
due to the unitary condition of the Schr$\ddot{o}$dinger equation.  
To our knowledge, it is of special interest to be dealt with the random potential described
by the form of fractional iterative map for the dynamical behavior of the quantum particle.  
Our extended random potential studied in this paper have not been fully explored up to now.
Our purpose in this study is to present the dynamical behavior of a quantum particle
through the fractional Schr$\ddot{o}$dinger equation, where
our potential is a time-dependent potential with the form 
of the fractional iterative map.

First of all the fractional Schr$\ddot{o}$dinger equation is represented in terms of
\begin{equation}
i\frac{\partial^{\tau}}{\partial t^{\tau}}  \Phi(x,t)  = -{\bigtriangledown}^{2\alpha} \Phi(x,t)+
W \rho(x,t) \Phi(x,t)
\label{at1}
\end{equation}
where $\Phi(x,t)$ is the wave function at displacment $x=x$ at time $t=t$.
Eq. $(1)$ is the fractional Schr$\ddot{o}$dinger equation reduced by dimensionless quantities such that
$ m=\frac{1}{2}$ and $ \frac{\hbar}{2m}=1 $.
The space scaling value $\tau$ and the time scaling value $\alpha$ are, repectively, the fractional values in the range
$ 0<\tau \le 1$ and $ \frac{1}{2} <\alpha \le 1$,
$W$ is the magnitute of a random potential, and $\rho(x,t)$ is a random function.
Let's introduce that a random function given by the form of the fractional iterative map $[18]$ at each displacement
is represented as follows: 
\begin{eqnarray}
\label{bs2}
{\rho \!(t)\!}=\! \left\{\!
\begin{array}{cc}
\gamma\exp [-\beta (-\log \rho(t-1))^{\eta}][1-\exp [-\beta (-\log \rho(t-1))^{\eta}]]
& \! \mbox{for} \! 0\! <\! \rho(t) \le\! 1 \, \\
-\gamma\exp [-\beta (-\log \rho(t-1))^{\eta}][1-\exp [-\beta (-\log \rho(t-1))^{\eta}]]
&\! \mbox{ for }\! -1\! \le \! \rho(t) \! \le\! 0 , 
\end{array} 
\right.
\label{bs2}
\end{eqnarray}
where $\beta$ and $\gamma$ are control parameters of the fractional iterative map,
and the ensemble-averaged values of a random function are
\begin{equation}
\overline{\rho(x,t)} \ = \ 0 \,
\label{ca3}
\end{equation}
and
\begin{equation}
\overline{\rho(x,t) \ \rho (x',t')} \ = \ \rho^2 (t) \ \delta_{x x^{'}} \  \delta (t-t^{'}). \
\label{db4}
\end{equation}
We assume the initial condition that the quantum particle is distributed by a Gaussian packet
centered at displacment $x=0$ at time $t=0$ with a width $\sigma$ $=$ $10$: 
\begin{equation}
 \Phi(x,t) = \frac{1}{\surd{2\pi {\sigma}^2}} \ e^{-x^2 / {2{\sigma}^2}}.
\label{ef5}
\end{equation}
Since the derivative of order $2\alpha$ for the wave function $\Phi(x,t)$ calculated by Crank-Nicolson method $[19]$
can be written in terms of
\begin{equation}
{\bigtriangledown}^{2\alpha} \Phi(x,t) = \frac{\Phi(x+1,t)-2\Phi(x,t)
+\Phi(x-1,t)}{{\Delta} x^{2\alpha}},
\label{fg6}
\end{equation}	
Eq.$(1)$ can be derived as
\begin{eqnarray}
\Phi(x,t+1)-\Phi(x,t) & = & \frac{i{\Delta} t^{\tau}}{2} \lbrace \lbrack 
{\bigtriangledown}^{2\alpha} \Phi(x,t+1) \nonumber \\
&& +{\bigtriangledown}^{2\alpha} \Phi(x,t) \rbrack
-\frac{W}{2} \lbrack \rho(x,t+1)  \nonumber \\
&& +\rho(x,t) \rbrack \lbrack \Phi(x,t+1)+\Phi(x,t) \rbrack \rbrace .  
\label{gh7}
\end{eqnarray}
Hence the quantum expectation values $[20]$ are defined by 
\begin{equation}
\langle x(t) \rangle = \langle \Phi(x,t) |x| \Phi(x,t)  \rangle ,
\label{hi8}
\end{equation}	
\begin{equation}
\langle p(t) \rangle =-\frac{i}{2} \langle \Phi(x,t) | \lbrack x,H \rbrack | \Phi(x,t)  \rangle ,
\label{ij9}
\end{equation}	
and
\begin{equation}
\langle f(t) \rangle =-\frac{W}{2} \langle \Phi(x,t) 
| \rho(x+1,t)- \rho(x-1,t) | \Phi(x,t)  \rangle .
\label{jk10}
\end{equation}
Here $\lbrack x,H \rbrack $ denotes the commutator of position operator $ x $ and Hamiltonian 
operator $ H $.
In our scheme, we will calculate numerically three quantum expectation values $ \overline {{\langle x(t) \rangle}^2}$, 
$ \overline {{\langle p(t) \rangle}^2}$, and $ \overline {{\langle f(t) \rangle}^2}$ from Eq.$(8)-(10)$,
and our numerical result that we have obtained will be also compared with that of other random potential.

For the sake of simplicity,  
we conveniently restrict ourselves to the random function, i.e. the form of fractional iterative map,
for three cases with $\eta$ $=$ $1.6$, $1.8$, and $2.0$.
In Eq. $(2)$, we focus on a random function of the fractional iterative map with control parameters $\beta=0.2$ and $\gamma=3.78$.
After finding the wave function $\Phi(x,t)$ in the special case of $\tau=1$ and $\alpha=1$, 
we discuss numerically the diffusive dynamical behavior of the quantum particle from quantum expectation values.
The quantum expectation values $ \overline {{\langle x(t) \rangle}^2}$, 
$ \overline {{\langle p(t) \rangle}^2}$, and $ \overline {{\langle f(t) \rangle}^2}$ 
ensemble-averaged over $2 \times 10^3$ configurations are exactly
found for fixed fractional value $\eta$ and the magnitude of a random potential $W$ $=$ $2,5,10$, and $15$,
and the result of these calculations is summarized in Table $1$.
Furthermore,
Fig. $1$ depicts $ \overline {{\langle x(t) \rangle}^2}$ as a function of time $t$,
while the plots of $ \overline {{\langle p(t) \rangle}^2}$ and $ \overline {{\langle f(t) \rangle}^2}$ 
are shown in Figs. $2$ and $3$, respectively.  
In the long time limit, it is obtained from Table $1$ that 
 $ \overline {{\langle x(t) \rangle}^2}$ $\propto$ $t^{\delta}$ with $\delta$ $=$ $0.96 \pm 0.02$, 
and that $ \overline {{\langle p(t) \rangle}^2}$ $\propto$ $t^{-\epsilon}$ and 
$ \overline {{\langle f(t) \rangle}^2}$ $\propto$ 
$t^{-\theta}$, where $\epsilon$ $=$ $0.98 \pm 0.02$ and $\theta$ $=$ $ 0.51 \pm 0.01$.

In conclusion, we have investigated 
the dynamical behavior for a fractional quantum Brownian particle
under a random potential of the fractional iterative
map on a one-dimensional lattice, and the scaling exponents 
for the quantum expectation value obtained
have also discussed at three fractional values $\eta$.
It has until now been well-known that $ \overline {{\langle x(t) \rangle}^2}$ $\propto$ $t$ $[10]$, 
if both the momentum and the random
potential are given by the white noise. Moreover, since the expectation value 
$ \overline {{\langle x(t) \rangle}^2}$
is proportional to $t^{3}$ for both the force and the random potential described by the white noise,
this case has been known to be similar to that of a classical random walker with no dissipation $[21]$.
In the large time limit, it has been reported $[6]$ that 
the scaling exponents are $\delta$ $=$ $0.5 \sim 0.6$ and 
$\epsilon$ $=$ $\theta$ $=$ $0.5$ if a random potential has the pattern of the white noise.
We obtain that our scaling exponents $\delta$ and $\epsilon$ take larger value than those of other
result for a quantum particle characterized by the subdiffusive motion $[9,17]$.
Our case would be considered to be consistent with the normal diffusion because of the feature of chaotic orbital density
of the fractional iterative map. 
It is also important to notice that our result for $ \overline {{\langle f(t) \rangle}^2}$ is in agreement 
with previous estimates $[6]$.
However, both cases can be recongnized as resulting from the continuous spread of the
quantum wave packet.
In future, we will be treated extensively with the dynamical behavior for a fractional quantum Brownian particle
under a random potential of the fractional iterative map, from the fractional Schr$\ddot{o}$dinger equation
at arbitrary fractional values $\tau$ and $\alpha$.
It is hoped that further detailed investigation for the fractional quantum Brownian motion will present
analytically and numerically elsewhere.\\

\vspace {5mm}
\newpage
%
\begin{center}
{\bf FIGURE  CAPTIONS}
\end{center}

\vspace {10mm}

\noindent
Figure $1$.  Plot of the quantum expectation value $ \overline {{\langle x(t) \rangle}^2}$ as a function of the time $t$
for a random function with $\eta=1.6$ for $W$ $=$ $2$, $5$, $10$, and $15$, where $\delta=0.94 \sim 0.98$.

\vspace {5mm}

\noindent
Figure $2$.  Plot of the quantum expectation value $ \overline {{\langle p(t) \rangle}^2}$ as a function of the time $t$
for a random function with $\eta=1.8$ for $W$ $=$ $2$, $5$, $10$, and $15$. The four lines have a near unity slope 
of $\epsilon=0.96 \sim 1.00$ which is consistent with the normal diffusion.

\vspace {5mm}

\noindent
Figure $3$.  Plot of the quantum expectation value $ \overline {{\langle f(t) \rangle}^2}$  as a function of the time $t$
for a random function with $\eta=2.0$ for $W$ $=$ $2$, $5$, $10$, and $15$, where $\theta=0.49 \sim 0.54$.

\vspace {15mm}

\begin{center}
{\bf TABLE  CAPTIONS}
\end{center}
\vspace {5mm}

\noindent
Table $1$. Summary of values of the scaling exponents $\delta$, $\epsilon$, and $\theta$
in the case of three fractional values $\eta$.

\vspace {3mm}

\begin{tabular}{|l|llrr|}  \hline
$                     $ & $W=2    $ & $W=5   $ & $W=10  $ & $W=15  $   \\ \hline
$\delta$   $(\eta=1.6)$ & $0.96   $ & $0.98  $ & $0.94  $ & $0.94  $   \\
$\epsilon$ $(\eta=1.6)$ & $1.00   $ & $1.00  $ & $0.94  $ & $0.97  $   \\
$\theta$   $(\eta=1.6)$ & $0.50   $ & $0.52  $ & $0.51  $ & $0.52  $   \\ \hline
$\delta$   $(\eta=1.8)$ & $0.97   $ & $1.01  $ & $0.95  $ & $0.91  $   \\
$\epsilon$ $(\eta=1.8)$ & $1.00   $ & $0.96  $ & $0.96  $ & $0.96  $   \\
$\theta$   $(\eta=1.8)$ & $0.50   $ & $0.51  $ & $0.51  $ & $0.50  $   \\ \hline
$\delta$   $(\eta=2.0)$ & $0.95   $ & $0.97  $ & $1.00  $ & $1.00  $   \\
$\epsilon$ $(\eta=2.0)$ & $1.00   $ & $0.98  $ & $0.96  $ & $1.00  $   \\
$\theta$   $(\eta=2.0)$ & $0.49   $ & $0.51  $ & $0.53  $ & $0.54  $   \\ \hline
\end{tabular}

\newpage
\begin{thebibliography}{110}
%
\bibitem{La1} P. Langevin, Comptes Rend Acad. Sci. Paris {\bf 146}, 530 (1908);
A. Einstein, $Theory$ $of$ $the$ $Brownian$ $Motion$ (Dover, 1956); E. Nelson,
$Dynamical$ $Theories$ $of$ $Brownian$ $Motion$ (Princeton Univ. Press, 1967).
\bibitem{Ga2} C. W. Gardner, $Handbook$ $of$ $Stochastic$ $Methods$, $2$nd ed. (Springer-Verlag, Berlin, 1985).
\bibitem{Ku3} R. Kubo, Rep. Prog. Phys. {\bf 29}, 255 (1966).
\bibitem{Mo4} H. Mori, Prog. Theor. Phys. {\bf 33}, 423 (1965); {\bf 34}, 399 (1965).
\bibitem{Le5} M. H. Lee, Phys. Rev. B{\bf 26}, 2547 (1982); Phys. Rev. Lett. {\bf 49}, 1072 (1982).
\bibitem{Sh6} P. Sheng and Z. Q. Zhang, Phys. Rev. B{\bf 48}, 609 (1993);
A. A. Ovchinnikov and N. S. Erikhman, Sov. Phys. JETP {\bf 40}, 733 (1975).
\bibitem{Ma7} A. Madhukar and W. Post, Phys. Rev. Lett. {\bf 39}, 1424 (1977).
\bibitem{Fe8} S. Feng, L. Glolubovic and H. Sphohn, M. J. Feigenbaum, Phys. Rev. Lett. {\bf 65}, 1028 (1990).
\bibitem{Me9} E. Medina, M. Kardar and H. Sphohn, Phys. Rev. Lett. {\bf 66}, 2176 (1991).
\bibitem{Sa10} L. Sand and M. Kardar, Phys. Rev. A{\bf 45}, 8859 (1992).
%
\bibitem{SC11} H. Scher and E. W. Montroll, Phys. Rev. B {\bf 12}, 2455 (1975).
\bibitem{Am12} F. Amblard, A. C. Maggs, B. Yurke, A. Pargellis and S. Leibler,
Phys. Rev. Lett. {\bf 77}, 4470 (1996).
\bibitem{HA13} S. Havlin and D. Ben-Avraham, Adv. in Phys. {\bf 36}, 695 (1987).
\bibitem{SO14} T. H. Solomon, E. R. Weeks and H. Swinney, Phys. Rev. Lett.
{\bf 71}, 3975 (1993).
\bibitem{SHL15} M. F. Schlesinger, B. J. West and J. Klafter, Phys. Rev. Lett.
{\bf 58}, 1100 (1987).
\bibitem{HYK16} B. I. Henry, M. K. Yum, Y. S. Kong, J. S. Choi and Kyungsik Kim, 
Chaos, Solitons and Fractals (to be published).
%
\bibitem{Bo17}J. P. Bouchard, D. Touati and D. Sornette, Phy. Rev. Lett. {\bf 16}, 2115 (1991).
\bibitem{Kim18}Kyungsik Kim, G. H. Kim, J. R. Lee, J. S. Choi, Y. S. Kong, B. I. Henry,
M. K. Yum and T. Odagaki, Fractals(to be published).
\bibitem{Sm19}G. D. Smith, $Numerical$ $Solution$ $of$ $Partial$ $Differential$ $Equation$ (Clarendon, Oxford, 1985) 
\bibitem{Sc20}L. I. Schiff, $Quantum$ $Mechanics$, 3rd ed. (McGraw-Hill, New York, 1968)
\bibitem{Uh21}G. E. Uhlenbeck and L. S. Ornstein, Phy. Rev. {\bf 36}, 823 (1930).
%
\end{thebibliography}
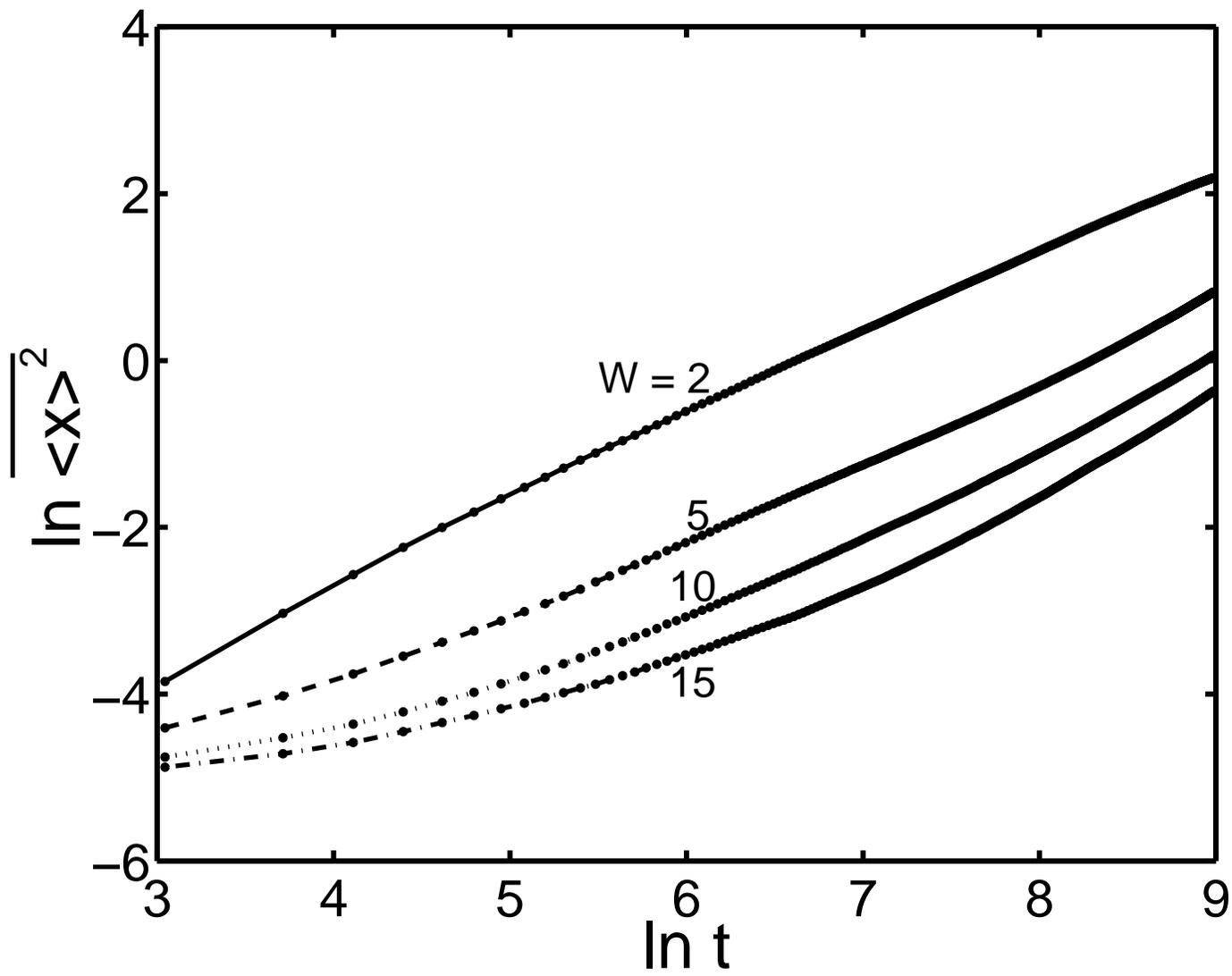
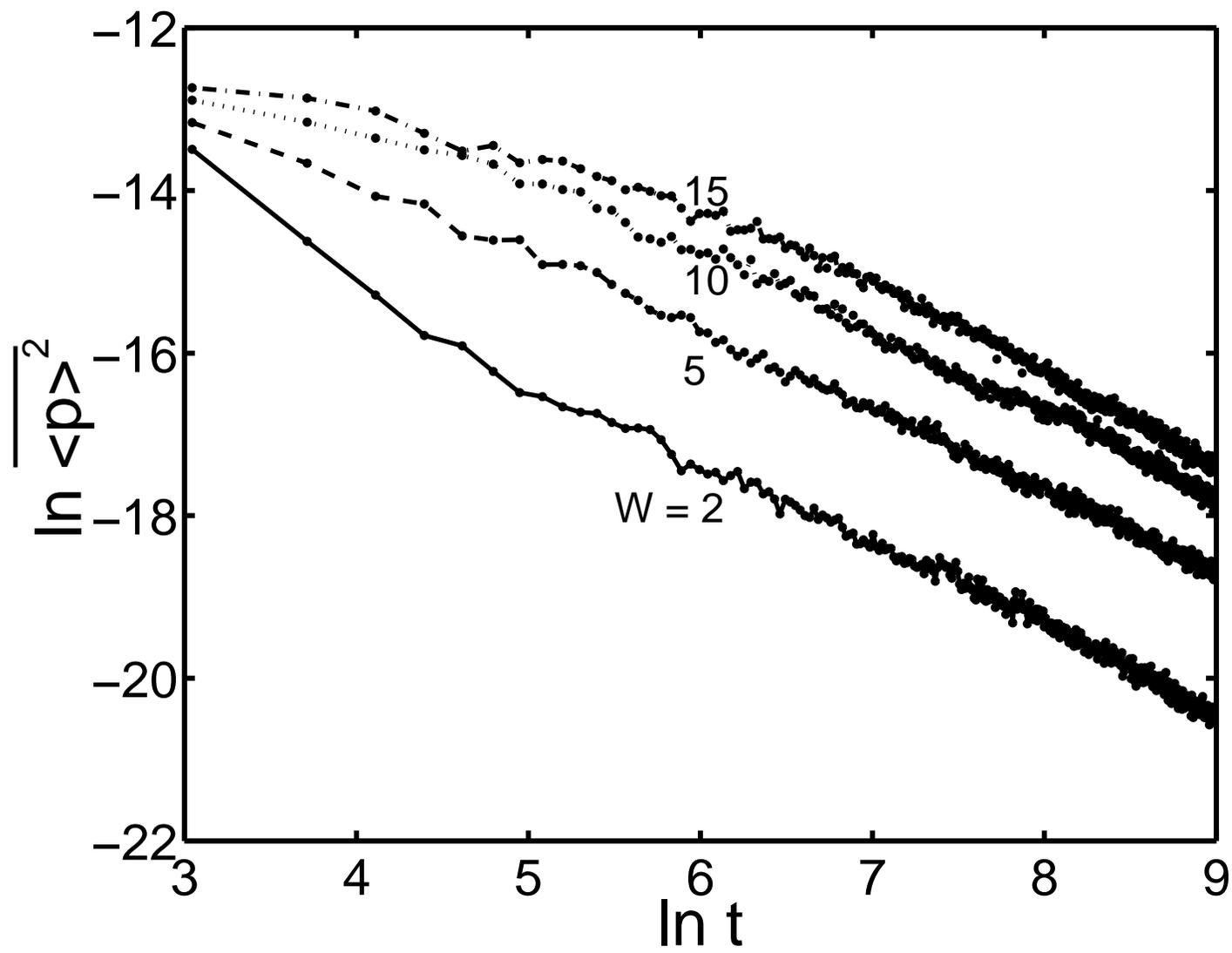
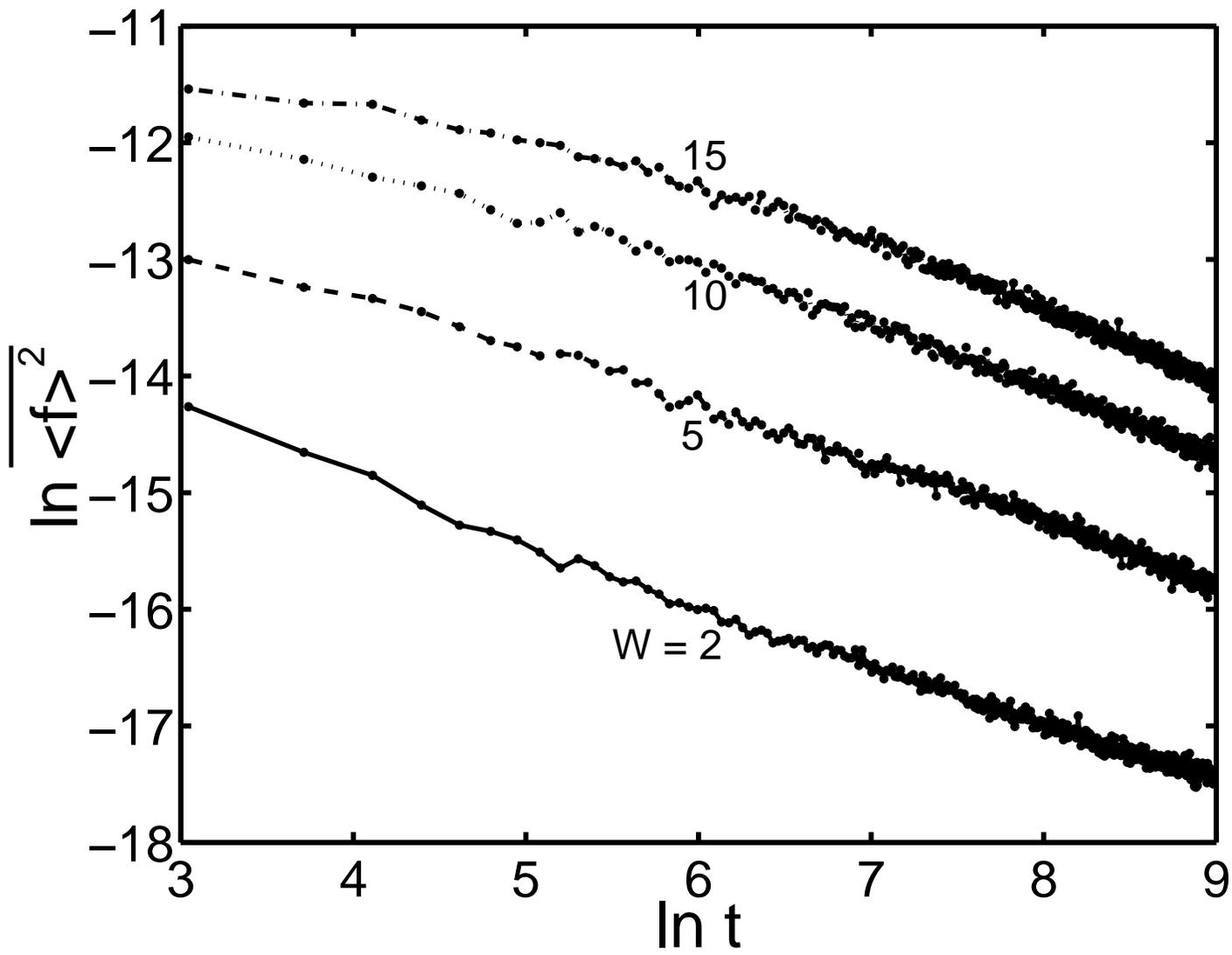
\end{document}